\documentclass[useAMS,usenatbib]{mn2e}
\usepackage{graphicx}
 
\newcommand{\oversim}[2]{\protect{\mbox{\lower0.5ex\vbox{%
  \baselineskip=0pt\lineskip=0.2ex
  \ialign{$\mathsurround=0pt #1\hfil##\hfil$\crcr#2\crcr\sim\crcr}}}}}
\newcommand{\simgreat}{\mbox{$\,\mathrel{\mathpalette\oversim>}\,$}} 
\newcommand{\simless} {\mbox{$\,\mathrel{\mathpalette\oversim<}\,$}} 

\title[Star-cluster formation]
{Implications for the
formation of star clusters from extra-galactic star-formation rates} 
\author[C. Weidner, P. Kroupa and S.S. Larsen]{C. Weidner$^{1,2}$\thanks{e-mail:
weidner@astrophysik.uni-kiel.de}, P. Kroupa$^{1,2}$\thanks{e-mail:
  kroupa@astrophysik.uni-kiel.de},\thanks{Heisenberg Fellow} and
  S.S. Larsen$^{3}$\thanks{e-mail: slarsen@eso.org}\\
$^{1}$Institut f\"ur Theoretische Physik und Astrophysik, 
              Universit\"at Kiel, 24098 Kiel, Germany\\
$^{2}$Sternwarte der Universit\"at Bonn, 53121 Bonn, Germany\\
$^{3}$European Southern Observatory, 85748 Garching, Germany}
\begin{document}

\date{Accepted . Received ; in original form }

\pagerange{\pageref{firstpage}--\pageref{lastpage}} \pubyear{2003}

\maketitle

\label{firstpage}

\begin{abstract}
Observations indicate that young massive star clusters in spiral and
dwarf galaxies follow a relation between luminosity of the brightest
young cluster and the star-formation rate (SFR) of the host galaxy, in
the sense that higher SFRs lead to the formation of brighter
clusters. Assuming that the empirical relation between maximum cluster
luminosity and SFR reflects an underlying similar relation between
maximum cluster mass ($M_{\rm ecl,max}$) and SFR, we compare the
resulting $SFR(M_{\rm ecl,max})$ relation with different theoretical models.
The empirical correlation is found to suggest that
individual star clusters form on a free-fall time-scale with their
pre-cluster molecular-cloud-core radii typically being a few~pc
independent of mass. The cloud cores contract by factors of~5 to
10~while building-up the embedded cluster. A theoretical
$SFR(M_{\rm ecl,max})$ relation in very good agreement with the
empirical correlation is obtained if the cluster mass function of
a young population has a Salpeter exponent $\beta \approx 2.35$
  and if this cluster population forms
within a characteristic time-scale of a few-$10$~Myr.  This short
time-scale can be understood if the inter-stellar medium is
pressurised thus precipitating rapid local fragmentation and collapse
on a galactic scale. Such triggered star formation on a galactic scale
is observed to occur in interacting galaxies. With a global SFR of
$3-5\,M_\odot$/yr the Milky Way appears to lie on the empirical
$SFR(M_{\rm ecl,max})$ relation, given the recent detections of very
young clusters with masses near $10^5\,M_\odot$ in the Galactic disk.
The observed properties of the stellar population of very massive
young clusters suggests that there may exist a fundamental
maximum cluster mass, $10^6 < M_{\rm ecl,max*}/M_\odot < 10^7$.
\end{abstract}


\section{Introduction}
In a series of publications Larsen \citep{Lar00, Lar01, Lar02} and
\citet{LarRi00} examined star cluster populations of 37 spiral and
dwarf galaxies and compared the derived properties with overall
attributes of the host galaxy. For this work they used archive HST
data, own observations and 
literature data. They showed that cluster luminosity functions (LFs)
are very similar for a variety of galaxies. They also found that the
V-band luminosity of the brightest cluster, $M_{\rm V}$, correlates
with the global star-formation rate, SFR, but it is unclear if this
correlation is of physical or statistical nature.  According to the
statistical explanation there is a larger probability of sampling more
luminous clusters from a universal cluster LF when the SFR is higher
\citep{Lar02, BiHuEl02}.

\citet{Lar01} concluded that all types of star clusters form according
to a similar formation process which operates with different
masses. Smaller clusters dissolve fast through dynamical effects (gas
expulsion, stellar-dynamical heating, galactic tidal field) and only
massive clusters survive for a significant fraction of a Hubble
time \citep{Ves98,FZ01,BaMa03}. The notion is that virtually all stars
form in clusters \citep{KB02,LaLa03}, and that a star-formation ``epoch''
produces a population of clusters ranging from about $5\,M_{\odot}$
(Taurus-Auriga-like pre-main sequence stellar groups) up to the
heaviest star cluster which may have a mass approaching
$10^6~M_\odot$. The time-scale over which such a cluster population
emerges within a galaxy defines its momentary SFR.

The aim of this contribution is to investigate if the empirical
$M_V(SFR)$ relation may be understood to be a result of physical
processes.  In \S~\ref{obs} the observational data concerning the
correlation between the SFR and $M_V$ of the brightest star cluster
are presented, and the empirical and physical models describing this
correlation are elaborated in \S~\ref{equ}.  \S~\ref{condis} contains
the discussion and conclusion.


\section{The observational data}
\label{obs}
Based on various observational results \citet{Lar02} concludes that
star clusters form under the same basic physical processes, and that
the so-called super-clusters are just the young and massive upper end
of the distribution. We firstly derive from this observational
material a correlation between the absolute magnitude of the brightest
cluster and the star-formation rate of the host galaxy.

\begin{figure}
\begin{center}
\vspace*{-1cm}
\includegraphics[width=8cm]{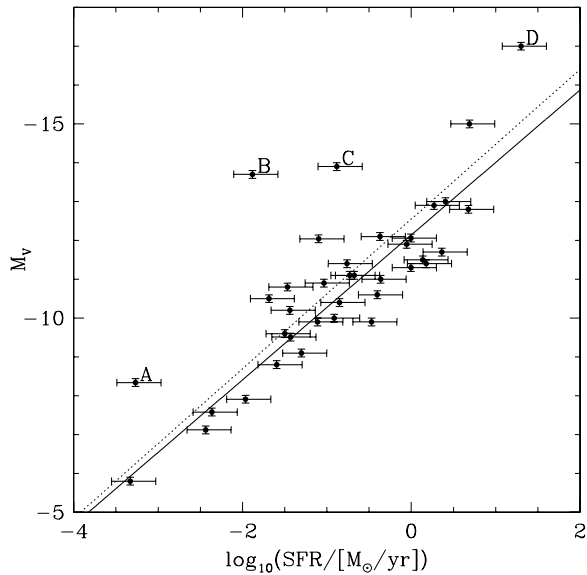}
\vspace*{-2cm}
\caption{Observational data from \citet{Lar02} for absolute magnitude
  of the brightest cluster versus global star-formation rate for 37
  disk and dwarf galaxies. The solid line illustrates a linear
  regression fit to the data excluding the four data points (A, DDO
  165), (B, NGC 1705), (C, NGC 1569) and (D, NGC 7252) while the
  dotted line includes all points.}
\label{fig1}
\end{center}
\end{figure}

Including in Fig.\ref{fig1} all data-points presented by
\citet{Lar01,Lar02} the following 
equation (\ref{fitall}) emerges from a 2-dimensional linear least square
fit
\begin{equation}\label{fitall}{\rm M}_{V} = -1.93 (\pm 0.06) \times \log{SFR} -
  12.55 (\pm 0.07)
\end{equation}
with a reduced $\chi^2_{\rm red}$ of about 17.
Excluding four points
\citep[A, B, C and D, see][]{Lar02} that
lie far above this first fit leads to
\begin{equation}\label{fit}{\rm M}_{V} = -1.87 (\pm 0.06) \times \log{SFR} -
  12.14 (\pm 0.07)
\end{equation}
with a reduced $\chi^2_{\rm red}$ of about 6. Both fits are shown in
Fig.~\ref{fig1}. For the magnitude ${\rm M}_{V}$ the {\it formal}
error is based on photon statistics, and is always very small
(especially since these are the brightest clusters in the galaxies),
usually 0.01 mag or less. Most of the errors are systematic, due to
uncertain aperture corrections, contamination within the photometric
aperture by other objects and are typically 0.1 mag.
The SFRs are derived from IR-fluxes puplished in the IRAS catalog
which lists typical errors of 15\%. However, a major source of
uncertainty in the derived SFRs lies in the FIR luminosity vs SFR
calibration, for which \citet{BuXu96} quote a typical error of
+100\%/-40\%.

Inverting eq.~\ref{fit} reveals,
\begin{equation}\label{invsfit}
\log{SFR} = -0.54 (\pm 0.02) \times {\rm M}_{V} - 6.51 (\pm 0.26),
\end{equation}
while a fit to the inverted data ($SFR$ vs ${\rm M}_V$) gives
\begin{equation}\label{invfit}
\log{SFR} = -0.54 (\pm 0.02) \times {\rm M}_{V} - 6.51 (\pm 0.19),
\end{equation}
with a reduced $\chi^2_{\rm red}$ of about 6. Both eqs.~\ref{invsfit}
and \ref{invfit} lead to essentially the same result thus nicely
demonstrating its robustness.

The exclusion of A, B, C and D is motivated by three of them being 
clusters in very sparse cluster systems in dwarf galaxies (DDO 165,
NGC 1705 and NGC 1569) dominated by a single brightest
member. Therefore the {\it present} SFR does not describe the rate
during the 
birth of these clusters. It has dropped to the shown values as no
further (massive) clusters are seen to be forming. This can be
understood as a general trend of aging after a star-formation epoch.
The underlying (observed) SFR has dropped while the clusters retain an
approximately constant luminosity for the first few million years
(Table~\ref{tab1}). The clusters therefore appear on the left in this
diagram (Fig.~\ref{fig1}) in dwarf galaxies in which SF proceeds in
bursts.  The cluster in NGC 7252 was excluded because this galaxy is a
several $10^8$~yr old merger in which the SFR was presumably much
higher when most of its clusters formed and in which the brightest
``cluster'' is probably an unresolved or already merged star-cluster
complex \citep{FeKr02a}, and thus not a true single cluster.

\section{The models}
\label{equ}
\subsection{Empirical model}
\label{empmod}
From the second fit to the observations (eq.~\ref{fit}) we derive an
empirical model for the dependence of the mass of the heaviest cluster
on the underlying star-formation rate of the host galaxy. With the
use of the mass-to-light-ratio, $k_{\rm ML}$, the magnitude (${\rm
M}_{\rm V}$) can be converted to a mass ($M_{\rm ecl,max}$),
\begin{equation}
\label{Mv}
{\rm M}_{\rm V} = 4.79 - 2.5 \cdot \log{\frac{M_{\rm ecl,max}}{k_{\rm ML}}},
\end{equation}
where $M_{\rm ecl,max}$ is the stellar mass in the cluster.  The
mass-to-light ratios in Table~\ref{tab1} are derived from
\citet[][fig.~7]{Smith01}. The age spread between 6.0 and 8.0 (in
logarithmic units) is used to estimate the mass errors for the
individual clusters in the Larsen data set plotted in
Fig.~\ref{fig4b} below.
\begin{table}
\begin{center}
\caption{\label{tab1} Mass-to-light ratios.}
\begin{tabular}{cc}
\hline
$k_{\rm ML}$&log age [yr]\\
\hline
0.0144&6.0 $< t <$ 6.8\\
0.0092&7.0\\
0.1456&8.0\\
\hline
\end{tabular}
\end{center}
\end{table}

Substituting eq.~\ref{fit} in \ref{Mv} gives,
\begin{equation}
\label{empMmax}
M_{\rm ecl,max} = k_{\rm ML} \cdot SFR^{0.75 (\pm 0.03)} \cdot 10^{6.77
  (\pm 0.02)}
\end{equation}

\noindent
and eq.~\ref{Mv} in \ref{invfit},
\begin{equation}
\label{empSFR}
SFR = \left(\frac{M_{\rm ecl,max}}{k_{\rm ML}}\right)^{1.34 (\pm 0.04)}
\cdot 10^{-9.07 (\pm 0.28)}
\end{equation}

\begin{figure}
\begin{center}
\vspace*{-1cm}
\includegraphics[width=8cm]{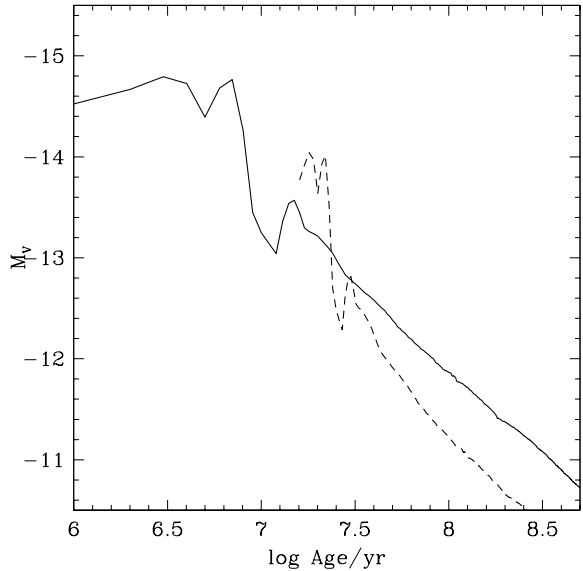}
\vspace*{-2cm}
\caption{Temporal evolution of the visual magnitude $M_{V}$ of a
$10^6\,M_{\odot}$ cluster (solid line) and a $5 \times
  10^5\,M_{\odot}$ cluster (dashed line) over 500 Myr with
  mass-to-light ratios
derived from a {\sc Starburst99} simulation \citep{Lei99}. In this
example the $5 \times 10^5\,M_{\odot}$ cluster forms when the
$10^6\,M_{\odot}$ cluster is $1.5 \times 10^7$ years old and appears brighter
than the more massive cluster for about 10 Myr.}
\label{fig:mlrevol}
\end{center}
\end{figure}

The question whether the brightest cluster observed is always the
heaviest is non-trivial to answer because for example a less-massive
but somewhat younger cluster may appear brighter than an older
more massive cluster, because the stellar population fades with
age. This does not always hold true for the very youngest phases,
where the clusters may briefly brighten somewhat due to the appearance
of red supergiant stars (Table~\ref{tab1}). We therefore
explore this problem with a rather simple model. Using three different
cluster formation rates (CFR; linear decreasing, linear increasing and
constant) a number of clusters is formed per time-step (1 Myr). Taking
a power-law CMF with an exponent $\beta=2$ (eq.~\ref{eq:cmf} in
\S~\ref{nolimit} below) cluster masses are allocated randomly by a
Monte-Carlo method. These clusters are then evolved using time dependent
mass-to-light ratios derived from a {\sc Starburst99} simulation
\citep{Lei99} for a Salpeter IMF ($\alpha = 2.35$) from
$0.18\,M_{\odot}$ to $120\,M_{\odot}$ for a $M_{\rm
  ecl}=10^6\,M_\odot$ cluster over 1 Gyr. The lower mass boundary is
chosen in order to have the same mass in stars in the cluster with the
Salpeter IMF as in a universal Kroupa IMF \citep{Krou01}.
The evolution of $M_V$ of a $M_{\rm ecl}=10^6\,M_\odot$ and a $M_{\rm
  ecl}=5 \times 10^5\,M_\odot$ cluster is shown in
Fig.~\ref{fig:mlrevol}. 

For the whole Monte-Carlo Simulation the heaviest cluster is also the
brightest for about 95\% of the time and for all three cases of the
CFR over the first 500 Myr. Therefore we
estimate an uncertainty of about 5\% on our assertion that the
brightest cluster in a population is also the most massive one. This
uncertainty can be neglected relatively to the larger uncertainties in
the cluster ages and therefore in the mass-to-light ratios.

\citet{Lar02} points out that a relation between the luminosity of the
brightest cluster, $M_{\rm V}$, and the total SFR arrived at by random
sampling from a power-law LF, given an area-normalised star formation
rate $\Sigma_{\rm SFR}$ and total galaxy size, reproduces the observed
correlation. The aim of this contribution is to investigate if the
correlation may be the result of physical processes. In essence, the
observed correlation is expected because in order to form a massive
cluster in a similar time-span a higher SFR is needed than for a
low-mass cluster.  In order to probe the physical background of the
empirical relation (eqs.~\ref{empMmax} and \ref{empSFR}) we calculate
a number of different models in \S~\ref{mwd} to \ref{tmm}.

\subsection{Local data model}
\label{mwd}
\begin{figure}
\begin{center}
\vspace*{-1cm}
\includegraphics[width=8cm]{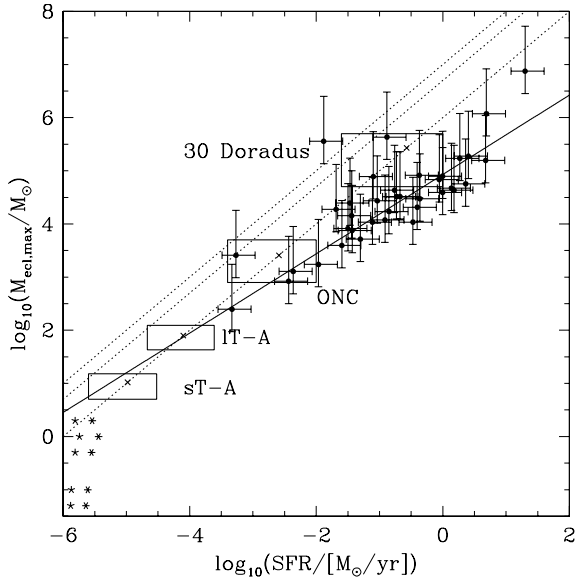}
\vspace*{-2cm}
\caption{Maximum cluster mass versus global star-formation rate (SFR),
both in logarithmic units. Filled dots are observations by Larsen with
error estimates (see~\S~\ref{empmod} for details) and the linear
regression fit is the solid line (eq.~\ref{empMmax}).
The crosses are different Galactic and extra-galactic clusters with
the SFR for each one obtained by simply dividing the mass by a
formation time of one Myr. The construction of the error boxes is
explained in~\S~\ref{mwd}. The individual clusters are: sT-A: small
sub-clumps in Taurus-Auriga, lT-A: the whole Taurus-Auriga
star-forming region, ONC: Orion Nebula cluster, 30 Doradus: the R136
cluster in the 30 Doradus star-forming region in the LMC. The dotted
lines show $M_{\rm ecl,max}(SFR)$ relations assuming a 1 (bottom), 5
and 10 (top) Myr formation time for individual clusters (ie.~not
cluster-systems). The two types of asterisks are single stars with
final main-sequence masses of $0.05, 0.1, .0.5, 1.0$ and
$2.0\,M_\odot$ (from bottom to top) after accretion of 90~pert cent
(on the left) and 99~per cent (on the right) of their mass, from
\citet{Wuch03}. }
\label{fig4b}
\end{center}
\end{figure}

In Fig.~\ref{fig4b} the data for individual clusters in the Milky Way
(Taurus-Auriga, Orion Nebula cluster) and in the LMC (R136, the
core of the 30 Doradus region) are compared with the
extragalactic cluster-system data. The 
data points (crosses) were calculated by dividing a mass estimate for
each cluster by a formation time of 1~My. This is a typical formation
time-scale as deduced from the ages of the stars in Taurus--Auriga,
the Orion Nebula cluster and in R136. The error for the mass scale is
constructed from different assumptions in the literature about the
number of stars in the cluster and with the use of different mean
masses as they vary in dependence of the used IMF and the maximal
possible stellar mass for the particular cluster. We thus have upper
and lower bounds on the cluster masses. By dividing the upper mass
over a formation time of 0.5 Myr and the lower mass over 2 Myr the
corresponding errors for the SFR are obtained. The data for these
assumptions are taken from \citet{Bric02}, \citet{Krou01},
\citet{Hart02}, \citet{Mass98}, \citet{Palla02} and \citet{Selm99}.

Fig.~\ref{fig4b} demonstrates that this simplest description already
leads to reasonable agreement with the observational data. That the
local individual cluster data are offset to lower SFRs from the
extragalactic data can be understood as being due to the observations
measuring the SFR for entire star-cluster populations rather than for
individual clusters and/or the formation time-scale to vary with
cluster mass. In \S~\ref{fft} the star-cluster formation time scale
(set here to be 1~Myr) is allowed to be the mass-dependent free-fall
time-scale.

\subsection{Free-fall model}
\label{fft}
\begin{figure}
\begin{center}
\vspace*{-1cm}
\includegraphics[width=8cm]{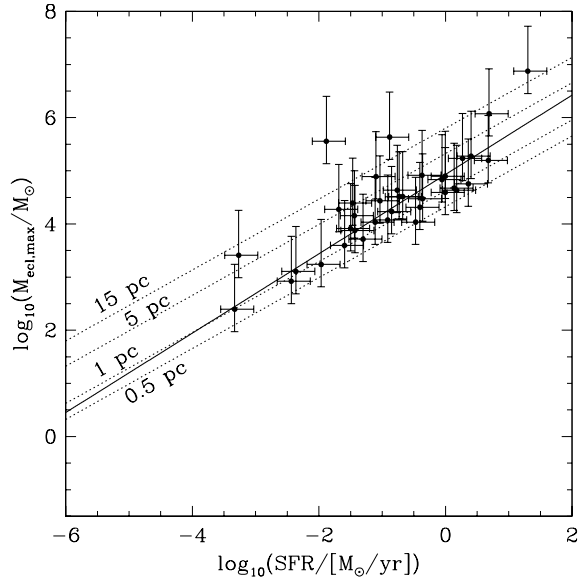}
\vspace*{-2cm}
\caption{As Fig.~\ref{fig4b} but the dotted lines show the model SFR
  needed to build a single cluster within one free-fall time
  (eq.~\ref{SFRtff}) for $R=0.5$, 1, 5 and 15~pc.}
\label{fig3b}
\end{center}
\end{figure}

Here the time scale for the formation of an individual star cluster is
the free-fall-time $t_{\rm ff}$ for a pre-cluster molecular cloud core
with radius $R$. This model is motivated by the insight by
\citet{Elme00} that star formation occurs on virtually every level,
from galactic scales over clusters to stars themselves, within one or
two crossing times. The SFR needed to build-up one (e.g. the most
massive) cluster in a free-fall time is
\begin{equation}\label{SFRt} SFR = \frac{M_{\rm ecl,max}}{t_{\rm ff}}.
\end{equation}
For the free-fall-time we take the dynamical time-scale \citep[for
  simliar considerations see e.g.][]{Elme00},
\begin{equation}\label{tff}t_{\rm ff} \approx \sqrt{\frac{R^{3}}{G
      \cdot M_{\rm st+g}}},
\end{equation}
where $M_{\rm st+g}$ is the total mass of the embedded cluster
including gas and stars. With a star-formation efficiency of 33\%
\citep{LaLa03,KrAaHu01} we
have $M_{\rm st+g} = 3 M_{\rm ecl,max}$. The combination of
\ref{SFRt} and \ref{tff} leads to
\begin{equation}\label{SFRtff}SFR = \sqrt{3}~M_{\rm ecl,max}^{3/2}
  \frac{\sqrt{\rm G}}{R^{3/2}},\end{equation} with G = $4.485 \cdot
10^{-3}\,{\rm pc^{3}}{M_{\odot}^{-1}\,{\rm Myr}^{-2}}$ being the
gravitational constant. Fig.~\ref{fig3b} shows this relation for
$R=0.5$, 1, 5 and 15~pc.

Thus a simple model based on the SFR required to form one cluster in a
free-fall time leads to a $M_{\rm ecl,max}(SFR)$ relation in good
agreement with the empirical relation provided the pre-cluster cloud
cores have radii of about 5~pc nearly independently of their
mass, because the correct relation ought to lie leftward of the
empirical data in Fig.~\ref{fig3b} since the empirical SFRs are for
entire cluster populations.  

The groups of pre-main sequence stars in Taurus-Auriga (a
few~$M_\odot$) have radii of about 0.5~pc \citep{Go93}. The about
1~Myr old Orion Nebula cluster (a few $1000\,M_\odot$) has a radius of
about 1~pc \citep{HiHa98} but it is most probably expanding owing to
gas expulsion \citep{KrAaHu01}.  The $1-3$~Myr old R136 in the LMC
($\approx 10^{4-5}\,M_\odot$), which today is seen to have a radius of
a few~pc \citep{Bran96}, is most likely also in a post-gas-expulsion
expansion phase. Also \citet{MA01} notes from his sample of 27 massive
($\simgreat 3 \times 10^4\,M_\odot$) and young ($< 20$~Myr) clusters that
those younger than about 7~Myr have radii of about 1~pc only. {\it
  Very young, still-embedded clusters appear to 
be very compact with radii of 0.5--1~pc, and the results from
Fig.~\ref{fig3b} can be taken to mean that they form in a free-fall
time if the pre-cluster cloud cores have radii of about~5~pc at the
onset of collapse}.  The build-up of the stellar population would
proceed while the density of the cloud core increases by a factor of
about $5^3$ to $10^3$, when the star-formation rate in the embedded
cluster probably peaks and declines rapidly thereafter as a result of
gas evacuation from accumulated outflows and/or the formation of the
massive stars that photo-ionise the cloud core \citep{MaMc00, TM02}.


\subsection{Total-mass model}
\label{tmm}
The above free-fall model quantifies the theoretical relation
for the case that the 
measurements only capture star-formation in the most massive
clusters in a galaxy. This can be considered to be a lower-bound on
the SFR. An upper bound is given by the rate with which all stars are
being formed, which means the total mass being converted to stars in a
given time interval. This total mass is the mass in the star-cluster
population and is the subject of this subsection, which begins by
assuming there exists no fundamental maximum star-cluster mass,
followed by an analysis in which a physical maximum cluster mass,
$M_{\rm ecl,max *}$, is incorporated.
 
\subsubsection{Without a physical maximum cluster mass}
\label{nolimit}
\begin{figure}
\begin{center}
\vspace*{-1cm}
\includegraphics[width=8cm]{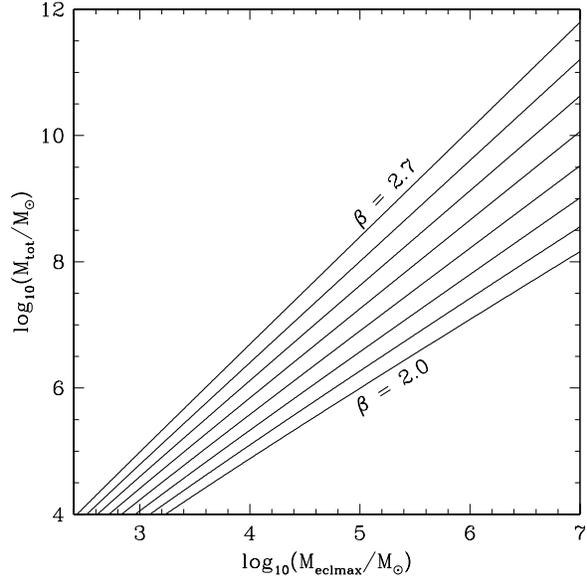}
\vspace*{-2cm}
\caption{The (logarithmic) total mass of a cluster system, $M_{\rm
tot}$, in dependence of the (logarithmic) maximal cluster mass,
$M_{\rm ecl,max}$, for different CMF power-law indices $\beta$ ($=2.0$
to 2.7, from bottom to top) }
\label{fig:totm}
\end{center}
\end{figure}

The aim is to estimate the SFR required to build a complete young
star-cluster population in one star-formation epoch such that it is
populated fully with masses ranging up to $M_{\rm
ecl,max}$. Observational surveys suggest the embedded-cluster mass
function (CMF) is a power-law,
\begin{equation} 
\label{eq:cmf}
\xi_{\rm ecl}(M_{\rm ecl}) = k_{\rm ecl} \cdot
  \left(\frac{M_{\rm ecl}}{M_{\rm ecl,max}}\right)^{-\beta},
\end{equation}
with $1.5\simless \beta\simless 2.5$ \citep[][and references
therein]{ElEf97,Krou02,KB02,LaLa03,KrWe03}.  For the total mass of a
population of young stellar clusters,
\begin{eqnarray} 
M_{\rm tot} = &\int_{M_{\rm ecl,min}}^{M_{\rm ecl,max}}M_{\rm ecl}
  \cdot \xi_{\rm ecl}(M_{\rm ecl})~dM_{\rm ecl} \nonumber\\
\label{Mint}
= & k_{\rm ecl} \cdot M_{\rm ecl,max}^{\beta} \cdot \int_{M_{\rm
  ecl,min}}^{M_{\rm ecl,max}}M_{\rm ecl}^{1-\beta}~dM_{\rm ecl},
\end{eqnarray}
where $M_{\rm ecl,max}$ is the mass of the heaviest cluster in the
population.  The normalisation constant $k_{\rm ecl}$ is determined by
stating that $M_{\rm ecl,max}$ is the single most massive cluster,
\begin{eqnarray} 1 =& \int_{M_{\rm ecl,max}}^{\infty}
  \xi_{\rm ecl}(M_{\rm ecl})~dM_{\rm ecl} \nonumber\\
\label{Nint}
= &k_{\rm ecl} \cdot M_{\rm ecl,max}^{\beta} \cdot \int_{M_{\rm
  ecl,max}}^{\infty} M_{\rm ecl}^{-\beta}~dM_{\rm ecl}.
\end{eqnarray}
With a CMF power-law index of $\beta = 2$ we get from eq.~\ref{Nint},
\begin{equation} \label{norm} 
k_{\rm ecl} = \frac{1}{M_{\rm ecl,max}}.
\end{equation}
Inserting this into eq.~\ref{Mint} (again with $\beta$ = 2),
\begin{equation} \label{mtot} M_{\rm tot} = M_{\rm ecl,max} \cdot
  (\ln{M_{\rm ecl,max}}-\ln{M_{\rm ecl,min}}).
\end{equation}
$M_{\rm ecl,min}$ is the minimal cluster mass which we take to be
$5\,M_{\odot}$ (a small Taurus-Auriga like group). For arbitrary
$\beta \ne 2$ eqs.~\ref{norm} and \ref{mtot} change to
\begin{equation} \label{normb} k_{\rm ecl} = \frac{\beta-1}{M_{\rm
      ecl,max}}
\end{equation}
and
\begin{equation} \label{mtotb} M_{\rm tot} = (\beta-1) \cdot M_{\rm
    ecl,max}^{\beta-1} \cdot \left(\frac{M_{\rm ecl,max}^{2-\beta}-M_{\rm
    ecl,min}^{2-\beta }}{2-\beta}\right).
\end{equation}
The resulting total mass, $M_{\rm tot}$, as a function of the maximal
cluster mass, $M_{\rm ecl,max}$, is shown in Fig.~\ref{fig:totm} for
different $\beta$. 

Given a SFR, a fully-populated CMF with total mass $M_{\rm tot}$ is
constructed in a time $\delta t$,
\begin{equation} \label{mtotc} M_{\rm tot} = SFR \cdot \delta t.
\end{equation}

\begin{figure}
\begin{center}
\vspace*{-1cm}
\includegraphics[width=8cm]{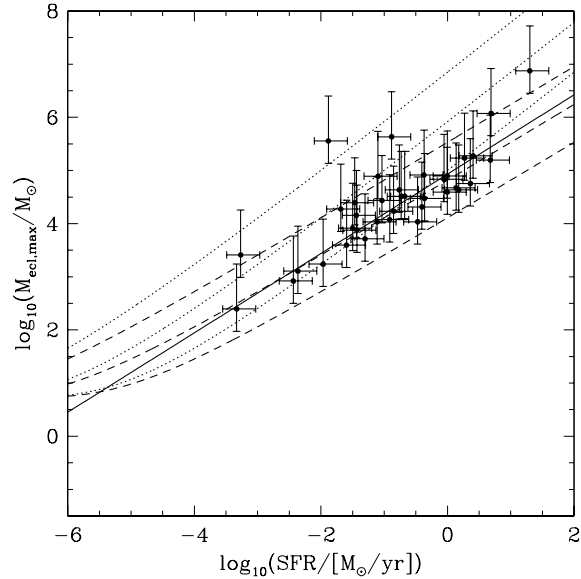}
\vspace*{-2cm}
\caption{Same as Fig.~\ref{fig4b}. However, here the theoretical
  relations 
(eqs.~\ref{mtot} and \ref{mtotc} or \ref{mtotb} and \ref{mtotc})
assume the entire young-cluster population forms in $\delta t=1$, 10
and 100~Myr (bottom to top). The CMF has $\beta=2$ (dotted
curves) or $\beta=2.4$ (dashed curves).}
\label{fig3a}
\end{center}
\end{figure}

Thus, dividing $M_{\rm tot}$ by different ad-hoc formation times,
$\delta t$, and using different maximal masses, $M_{\rm ecl,max}$,
results in a series of theoretical $M_{\rm ecl,max}(SFR)$ relations
which are shown in Fig.~\ref{fig3a}. 
{\it It thus appears that star-formation epochs with duration $\delta
t \approx 10$~Myr suffice for populating complete
cluster systems.}

\begin{figure}
\begin{center}
\vspace*{-1cm}
\includegraphics[width=8cm]{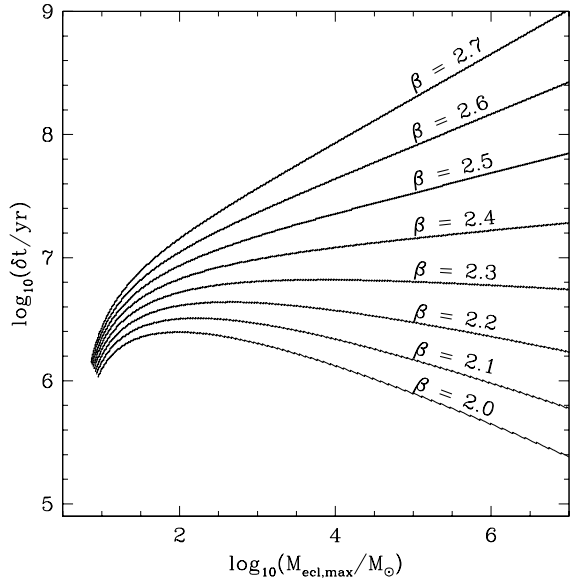}
\vspace*{-2cm}
\caption{Formation time scale (logarithmic years) of the cluster
system (eqs~\ref{dtb2} and ~\ref{dtc}) over maximal cluster mass (in
logarithmic units) for different slopes $\beta$ of the CMF assuming
the mass-to-light ratio of young clusters is $k_{\rm ML} = 0.0144$
(Table~\ref{tab1}). }
\label{figdt}
\end{center}
\end{figure}

The argumentation can now be inverted to better quantify the
time-scale required to build an entire young cluster population in a
star-formation epoch with a given SFR. For this purposed we employ the
empirical $SFR(M_{\rm ecl,max})$ relation. For $\beta = 2$,
\begin{equation} \label{dt} \delta t = \frac{M_{\rm tot}}{SFR} = \frac{M_{\rm
      ecl,max}}{SFR} \ln\left(\frac{M_{\rm ecl,max}}{M_{\rm ecl,min}}\right).
\end{equation}
Eq.~\ref{empSFR} can be re-written,
\begin{equation}
\label{empSFRb}
SFR = \left(\frac{M_{\rm ecl,max}}{k_{\rm ML}}\right)^{s}
    10^{-9.07}, 
\end{equation}
with $s$ = 1.34 being the exponent of this empirical $SFR(M_{\rm
ecl,max}$) relation. Combining eqs.~\ref{dt} and
\ref{empSFRb} we finally arrive at
\begin{equation}
\label{dtb2}
\delta t = M_{\rm ecl,max}^{1-s} \ln\left(\frac{M_{\rm
    ecl,max}}{M_{\rm ecl,min}}\right) k_{\rm ML}^{s} \cdot
    10^{9.07}~{\rm [yr]}. 
\end{equation}
For $\beta \ne 2$ we obtain instead,
\begin{eqnarray}
\delta t = & (\beta-1)~M_{\rm ecl,max}^{\beta-1-s}
\left(\frac{M_{\rm ecl,max}^{2-\beta}-M_{\rm
    ecl,min}^{2-\beta}}{2-\beta}\right) \times \nonumber\\
\label{dtc}
& k_{\rm ML}^{s} \cdot 10^{9.07}~{\rm [yr]}.
\end{eqnarray}
The cluster-system formation time scale, or the duration of the
star-formation ``epoch'', $\delta t$, is plotted in Fig.~\ref{figdt}
for different $M_{\rm ecl,max}$ -- and therefore different $M_{\rm
tot}$ -- and different CMF slopes $\beta$. For $\beta\simless2.4$ a
decreasing $\delta t$ for almost all masses is found which indicates
that the formation of the whole cluster system can be very rapid
($\simless 10$~Myr).

{\it We have thus found that the empirical $SFR(M_{\rm ecl,max})$
relation implies that more-massive cluster populations need a shorter
time to assemble than less massive populations, unless the embedded
cluster mass function is a power-law with an index of
$\beta\approx 2.35$, strikingly similar to the Salpeter index for
stars $(\alpha=2.35)$.} 
In this case the formation time becomes $\approx$ 10 Myr
  independent of the maximum cluster mass in the population.

Young populations of star clusters extend to 
super-star clusters mostly in galaxies that are being perturbed or
that are colliding.  The physics responsible for this can be sought in
the higher pressures in the inter-stellar medium as a result of the
squeezed or colliding galactic atmospheres \citep{ElEf97,
BeCo03}. When this occurs, massive molecular clouds rapidly build-up
and collapse locally but distributed throughout the galaxy. If two
disk galaxies collide face-on, star-formation occurs synchronised
throughout the disks, while edge-on encounters would lead to the
star-formation activity propagating through the disks with a velocity
of a few-100~pc/Myr (the relative encounter velocity) which amounts to
a synchronisation of star-formation activity throughout 10~kpc radii
disks to within a few-10~Myr. Just recently \citet{EnPlRoBl03} found
that for M33 the typical lifetimes of giant molecular clouds (GMC)
with masses ranging up to $7 \times 10^5 M_{\odot}$ are $10~{\rm
to}~20 \times 10^6$~yr, indicating a similar formation time for star
clusters born form these clouds. \citet{HBB01} deduce from
solar-neighbourhood clouds that their 
life-times are also comparable to the ages of the pre-main sequence
stars found within them, again suggesting that molecular clouds form
rapidly and are immediately dispersed again through the immediate
on-set of star-formation.

Notable is that $\beta\approx2.4$ gives a theoretical $M_{\rm
ecl,max}(SFR)$ relation with virtually the same slope as the empirical
relation (Fig.~\ref{fig3a}). {\it The implication would be that the
embedded-CMF is essentially a Salpeter power-law.} Also, in the
analysis above we neglected to take into account that once an embedded
cluster expels its residual gas it expands and loses typically $1/2$
to $2/3$ of its stars \citep{Krou02,KB02}. The
observed clusters with ages $>10\,$~Myr thus have masses
$(0.3-0.5)\times M_{\rm ecl}$. Taking this into account would shift
the theoretical relations downwards by at most~0.5 in log-mass which
would lead to an increase in $\delta t$ by a factor of a few.

\subsubsection{With a physical maximal cluster mass}
\label{lim}
The most massive ``clusters'' known, e.g. $\omega$~Cen \citep[a
  few~$10^6\,M_\odot$,][]{Gned02} or G1 \citep[$\approx 15\times
  10^6\,M_\odot$,][]{Meyl01} consist of complex stellar populations
with different 
metalicities and ages \citep{HiRi00}.  They are therefore not
single-metalicity, single-age populations that arise for a truly
spatially and temporarily localised star-cluster forming event, but
are probably related to dwarf galaxies that formed from a compact
population of clusters and with sufficient mass to retain their
interstellar medium for substantial times and/or capture field-stellar
populations and/or possibly re-accrete gas at a later time to form
additional stars \citep{Krou98, FeKr02b}. A fundamental, or physical
maximal star-cluster mass may therefore be postulated to exist on
empirical grounds in the range $10^6 \simless M_{\rm ecl,max*}/M_\odot
\simless 10^7$. In the following we explore the implications of such a
fundamental maximum cluster mass on the analysis presented in
\S~\ref{nolimit}.

\begin{figure}
\begin{center}
\vspace*{-1cm}
\includegraphics[width=8cm]{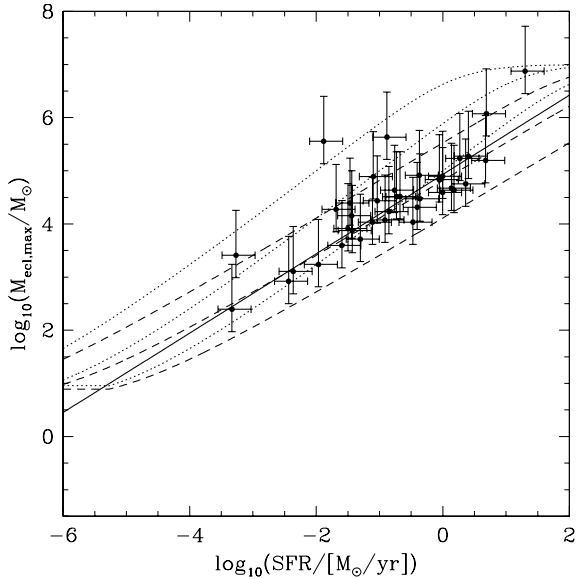}
\vspace*{-2cm}
\caption{As Fig.~\ref{fig3a} but for the case that there exists a
  fundamental maximum cluster mass $M_{\rm ecl,max*}=
  10^{7}\,M_{\odot}$.}
\label{fig3aa}
\end{center}
\end{figure}

For the following $M_{\rm ecl,max*}= 10^7\,M_{\odot}$ is adopted.
Eq.~\ref{Mint} remains unchanged while eq.~\ref{Nint} changes to
\begin{equation}
\label{Nintlim}
1 = k_{\rm ecl} \cdot M_{\rm ecl,max}^{\beta} \cdot \int_{M_{\rm
  ecl,max}}^{M_{\rm ecl,max*}} M_{\rm ecl}^{-\beta}~dM_{\rm ecl}.
\end{equation}
This can be evaluated for $\beta \ne 1$
\begin{equation}
k_{\rm ecl} = M_{\rm ecl,max}^{-\beta} \frac{1-\beta}{M_{\rm
    ecl,max*}^{1-\beta}-M_{\rm ecl,max}^{1-\beta}}.
\end{equation}
If $\beta=2$,
\begin{equation}
\label{Mtotlim}
M_{\rm tot} = - \frac{\ln{M_{\rm ecl,max}}-\ln{M_{\rm ecl,min}}}{M_{\rm
    ecl,max*}^{-1}-M_{\rm ecl,max}^{-1}},
\end{equation}
or for $\beta \ne 2$
\begin{equation}
\label{Mtotlimb}
M_{\rm tot} = \frac{1-\beta}{2-\beta} \frac{M_{\rm
    ecl,max}^{2-\beta}-M_{\rm ecl,min}^{2-\beta}}{M_{\rm
    ecl,max*}^{1-\beta}-M_{\rm ecl,max}^{1-\beta}}.
\end{equation}
For a fixed $M_{\rm ecl,max*}$ and a changing $M_{\rm tot}$ the upper
mass $M_{\rm ecl,max}$ for each cluster system can now be evaluated.
The resulting $SFR(M_{\rm ecl,max})$ models are plotted in
Fig.~\ref{fig3aa} for different formation-times of the entire cluster
population. The conclusions of the previous section do not change. 

\begin{figure}
\begin{center}
\vspace*{-1cm}
\includegraphics[width=8cm]{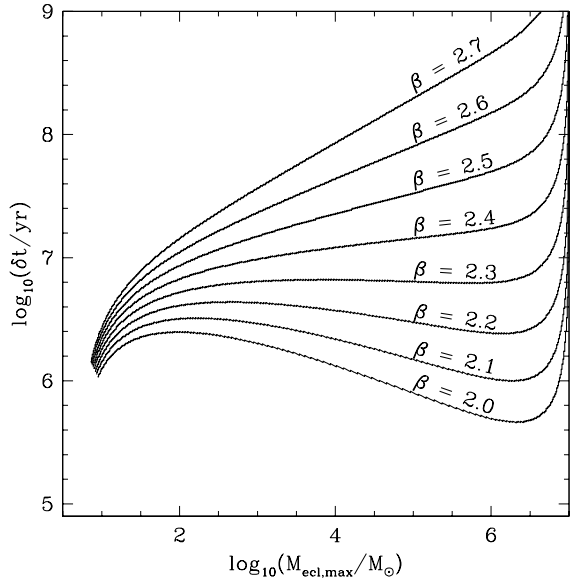}
\vspace*{-2cm}
\caption{As Fig.~\ref{figdt} but assuming the fundamental cluster mass
  limit is $M_{\rm ecl,max*}=10^7\,M_\odot$.}
\label{figdt2}
\end{center}
\end{figure}

Given the empirical $SFR(M_{\rm ecl,max})$ relation, the time-scale,
$\delta t$, needed to build-up a fully populated young star-cluster
population can be determined as in \S~\ref{nolimit}. The result is
shown in Fig.~\ref{figdt2}. Note that in both the limited ($M_{\rm
ecl,max*}=10^7\,M_\odot$, Fig.~\ref{figdt2}) and the unlimited
($M_{\rm ecl,max*}=\infty$, Fig.~\ref{figdt}) case it takes an
arbitrarily long time to sample the CMF arbitrarily close to $M_{\rm
ecl,max*}$.

\section{Discussion and Conclusions}
\label{condis}

Observations of young star-cluster systems in disk galaxies show that
there exists a correlation between the total SFR and the luminosity of
the brightest star-cluster in the young-cluster population. This can be
transformed to a SFR--heaviest-cluster-mass relation ($SFR(M_{\rm
ecl,max})$, eq.~\ref{empSFR}).

Very young star-clusters in the MW and the LMC that are deduced to
have formed within a few~Myr follow a similar $SFR(M_{\rm ecl,max})$
relation, although this ``local'' relation is somewhat steeper if it
is assumed that the formation time-scale of individual clusters is the
same in all cases ($\approx 1$~Myr, Fig.~\ref{fig4b}). Taking instead
the formation-time-scale to be the free-fall time of the pre-cluster
molecular cloud core the correct slope is obtained if the pre-cluster
cloud core radius is a few~pc independent of cluster mass
(Fig.~\ref{fig3b}).  This implies that the cluster-forming molecular
cloud cores may contract by a factor of~5 to~10 as the clusters form.
That the pre-cluster radii appear to 
not vary much with cluster mass
implies the pre-cluster cores to have increasing density with
increasing mass. Indeed, \citet{Lar03} finds young extra-galactic
clusters to have only a mild increase of effective radius with mass,
and embedded clusters from the local Milky Way also suggest the
cluster radii to be approximately independent of cluster mass
\citep{Krou02, KB02}.

A model according to which the total mass of the young-cluster
population, $M_{\rm tot}$, is assumed to be assembled in a
star-formation ``epoch'' with an a-priori unknown duration, $\delta
t$, gives the corresponding $SFR=M_{\rm tot}/\delta t$ and leads to
good agreement with the empirical $SFR(M_{\rm ecl,max})$ relation for
$1\simless \delta t/{\rm Myr}\simless 10$. A particularly good match
with the empirical relation results for $\delta t\approx {\rm
few}\times 10$~Myr and for a power-law CMF with $\beta\approx 2.35$. It
should be noticed that the slope of this CMF for stellar clusters is
virtually the same as for the Salpeter IMF ($\alpha=2.35$) which
applies for the early-type stars in these clusters.
Conversely, adopting the empirical $SFR(M_{\rm
ecl,max})$ relation, $\delta t$ can be calculated for different
young-cluster power-law mass functions with exponent $\beta$. We find
that $\delta t \simless {\rm few}\times 10$~Myr for
$\beta\simless2.4$. This value is nicely consistent with independent
observations. For example, \citet{HEDM03} find $2 < \beta < 2.4$ for a
sample of 939 LMC and SMC clusters after applying corrections for
redding, fading, evaporation and size-of-sample effects.

The same holds true if a fundamental maximum
star-cluster mass near $M_{\rm ecl,max*}=10^7\,M_\odot$ is introduced.
The existence of such a fundamental maximum cluster mass is supported
by ``clusters'' with $M \simgreat 5\times 10^6\,M_\odot$ having
complex stellar populations more reminiscent of dwarf galaxies that
cannot be the result of a truly single star-formation event.

The short time-span $\delta t \approx {\rm few} \times 10$ Myr for
completely-populating a CMF up to 
the maximum cluster mass of the population, $M_{\rm ecl,max}\le M_{\rm
ecl,max*}$, can be understood as being due to the high ambient
pressures in the inter-stellar medium needed to raise the global SFR
high enough for populous star-clusters to be able to emerge.  This
short time-scale, which we refer to as a star-formation ``epoch'',
does not preclude the star-formation activity in a galaxy to continue
for many ``epochs'', whereby each epoch may well be characterised by
different total young-star-cluster masses, $M_{\rm tot}$. According to
this notion, dwarf galaxies may experience unfinished ``epochs'', in
the sense that during the onset of an intense star-formation activity
that may be triggered through a tidal perturbation for example, the
ensuing feedback which may include galactic winds may momentarily
squelch further star-formation within the dwarf such that the cluster
system may not have sufficient time to completely populate the cluster
mass function. Squelching would typically occur once the most massive
cluster has formed. Dwarf galaxies would therefore deviate notably
from the $M_{\rm ecl,max}(SFR)$ relation (\S~\ref{obs}).

The conclusion is therefore that the observed $SFR(M_{\rm ecl,max})$
data can be understood as being a natural outcome of star formation in
clusters and that the SFR at a given epoch dictates the range of
star-cluster masses formed given a CMF that appears to be a Salpeter
power law. The associated formation time-scales are short being
consistent with the conjecture by \citet{Elme00} that star-formation
is a very quick process on all scales. Within about $10^7$~yr a
complete cluster system is build (Fig.~\ref{fig3a},~\S~\ref{fft}),
while individual clusters form on a time scale of $10^6$ years and
stars only in about $10^5$ years. Correspondingly,
  molecular-cloud life-times are short ($\approx {\rm few} \times 10$
Myr) - suporting the assertion by \citet{HBB01}.

Applying the empirical $SFR(M_{\rm ecl,max})$ relation to the MW which
has $SFR \approx 3-5\,M_\odot$/yr \citep{PrAu95} a
maximum cluster mass of about $10^5\,M_\odot$ is expected from
eq.~\ref{empMmax}. It is interesting that only recently have
\citet{AlHo03} revealed a very massive cluster in our Milky-way with
about 100 O stars (similar to R136 in the LMC). \citet{Kn00} notes
that the Cygnus~OB2 association contains $2600\pm400$ OB stars and
about 120~O~stars with a total mass of $(4 - 10) \times
10^4\,M_\odot$, and that this ``association'' may be a very young
globular-cluster-type object with a core radius of approximately 14~pc
within the MW disk at a distance of about 1.6~kpc from the Sun
\citep[but see][]{BiBoDu03}.  This object may be expanded after violent gas
expulsion \citep{BK03a, BK03b}. The MW therefore does not
appear to be unusual in its star-cluster production behaviour.

\section*{Acknowledgements}
This work has been funded by DFG grants KR1635/3 and KR1635/4.

\bsp

\label{lastpage}

\end{document}